Title: In the quest for a global (super)symmetry-breaking mechanism: an extension of the vibronic mixing process to superpartners
Author: Mladen Georgiev (Institute of Solid State Physics, Bulgarian Academy of Sciences, 1784 Sofia, Bulgaria)
Comments: 7 pdf pages
Subj-class: physics

We propose a global symmetry-breaking mechanism based on an extension of the vibronic (cooperative Jahn-Teller) effects (comprising electron-phonon superpartners) to generalize them into fermion-boson superpartners. Examples of superpairs are constructed by combining phonons, photons, π-mesons, etc. with electrons, holes, nucleons, etc. in condensed and nuclear matter. The obtained symmetry-breaking mechanism operating amidst the super-partner pairs may be just what is needed to justify the status of supersymmetry as a broken symmetry phenomenon. Following the extended vibronic mixing of fermion states by the appropriate boson superpartner, the latter propagates along a helical path away from the mixing site.

1. Introduction

In earlier preprints, we considered possible extensions of mechanisms we have become familiar with in condensed matter to nuclear matter providing corresponding interpretations to deformed nuclei [1], colossal molecular binding [2], and debated helical form of light propagation [3]. Now we aim at applying global extensions of well-known condensed matter effects, such as the cooperative Jahn-Teller or vibronic mixing of electron orbital states, to supersymmetry in order to see whether these do not meet the conditions for the expected supersymmetry-breaking mechanism [4]. It turns out that the natural classical path of a partner-boson following a mixing event is one along a helix [3]. Considering the importance of these conclusions, a more refined analysis is necessary and will be sought herein.

2. Hamiltonian for extended vibronic mixing

We consider an extended global Hamiltonian introducing ladder operators in $2^{nd}$ quantization, $a_{n\alpha s}$ and $a_{n\beta s}^{\dagger}$, etc. for fermions, $b_{ns}$ and $b_{ns}^{\dagger}$ for bosons. Their mixing Hamiltonian reads

$$H_{mix} = \sum_{n\alpha\beta s} g_{n\alpha\beta s} (b_{ns}^{\dagger} + b_{ns})(a_{n\alpha s}^{\dagger} a_{n\beta s} + a_{n\beta s}^{\dagger} a_{n\alpha s}) \qquad (1)$$

where $g_{n\alpha\beta s} = G_{n\alpha\beta}(M_{n\alpha\beta s}\omega_{n\alpha\beta s}^{2}/h\nu_{n\alpha\beta s})^{-½}$ is the mixing constant, $M_{n\alpha\beta s}$ the reduced mass and $\nu_{n\alpha\beta s}$ the vibrational frequency of the coupled oscillator, $\alpha$ and $\beta$ are two orbital fermion bands, nearly-degenerate and of opposite parities, s is the spin variable, n is the site index. $a_{n\alpha s}^{\dagger}$ and $b_{ns}^{\dagger}$, etc. are superpartners. To obtain the complete global Hamiltonian, we add up the diagonal energy terms, fermion and boson, respectively:

$$H_{f} + H_{b} = \sum_{n\alpha s} E_{n\alpha s}\, a_{n\alpha s}^{\dagger} a_{n\alpha s} + \sum_{n\alpha s} (n + ½) h\nu_{n\alpha s}\, b_{ns}^{\dagger} b_{ns} \qquad (2)$$

as well as the fermion hopping (band) term

$$H_{fhop} = \sum_{n\alpha s} t_{n\alpha s}\, a_{n\alpha s}^{\dagger} (a_{n-1\alpha s} + a_{n+1\alpha s}) \qquad (3)$$

The complete Hamiltonian then obtains as

$$H = H_f + H_b + H_{fhop} + H_{mix} \qquad (4)$$

The philosophy of (1) is clear: it is composed of products of creation operators in the one fermion band and annihilation operators in the other one. These products are then factorized by boson creation (annihilation) operators to account for the appearance (disappearance) of a boson in the fermion band mixing process. The bands are narrow, all but single levels, which makes the theory easily adaptable to the latter case. Looking in retrospect, the fermion bands appear as a result of level broadening due to interparticle coupling. The additive character of the boson-operator dependent terms in (1) follows from the requirement that the classical boson coordinate, proportional to the ladder constants (before subjecting them to commutator relations to become operators), must be real.

Schrödinger's equation based on (4) has been solved using Merrifield's Ansatz for the field operator (wavefunction) yielding curious results for the vibronic polaron [5-7]. For instance, the average phonon coordinate of a semibound polaron was found to attain a peak value in the middle of the quadrant and valleys at its edges.

### 3. Adiabatic approximation for light fermions versus heavier bosons

### (example: electrons vs. phonons, vibronic polarons)

To see just what classical trajectory is associated with boson propagation before and after a mixing event, we introduce the boson coordinate $q_{\alpha\beta s} = (M_{n\alpha\beta s}\omega_{n\alpha\beta s}^2/h\nu_{n\alpha\beta s})^{-\frac{1}{2}} (b_{ns}^{\dagger} + b_{ns})$ (site index n may be dropped for convenience though implied in an isotropic medium) and apply the adiabatic theorem [8] whereby solving for the Schrödinger equation is split in two parts, electronic and vibronic. The former is dealt with as the kinetic energy terms, fermion and boson, are discarded, while the boson coordinate is assumed to be a c-number parameter. The electronic part is solved in a two-site approach leading to double-well adiabatic energies. The adiabatic and vibrational terms in (5) then play the role of vibronic potentials for (6). More specifically,

$$H_{fermion} = \sum_{\alpha\beta s} E_{\alpha s} a_{\alpha s}^{\dagger} a_{\alpha s} + \tfrac{1}{2}\sum_{\alpha\beta s} M_{\alpha\beta s}\omega_{\alpha\beta s}^2 q_{\alpha\beta s}^2 + \sum_{\alpha\beta s} G_{\alpha\beta s} q_{\alpha\beta s}(a_{\alpha s}^{\dagger} a_{\beta s} + a_{\beta s}^{\dagger} a_{\alpha s}) \qquad (5)$$

$$H_{vibron} = -(h^2/4\pi m)\sum_{i\alpha\beta s}(\partial^2/\partial_i q_{\alpha\beta s}) + \tfrac{1}{2}\sum_{\alpha\beta s} M_{\alpha\beta s}\omega_{\alpha\beta s}^2 q_{\alpha\beta s}^2 + \sum_{\alpha\beta s} G_{\alpha\beta s} q_{\alpha\beta s}(a_{\alpha s}^{\dagger} a_{\beta s} + a_{\beta s}^{\dagger} a_{\alpha s})$$

$$+ \sum_{\alpha\beta s} E_{\alpha s} a_{\alpha s}^{\dagger} a_{\alpha s} \qquad (6)$$

Under these conditions the mixing terms in (6) stimulate a free rotation around the central site with pre-given radius (off-center displacement) but in order to make the situation more realistic, we complement these terms by the addition of third-order terms in $q_{\alpha\beta s}$ such that the boson coordinate part of (5) and (6) becomes $G_{\alpha\beta s} q_{\alpha\beta s} + D_{\alpha\beta s} q_{\alpha\beta s}^3$ to substitute for the linear coordinate term $G_{\alpha\beta s} q_{\alpha\beta s}$ [9]. Now the off-diagonal terms in the vibronic equation yield fourth order in $q_{\alpha\beta s}$ (higher terms neglected) to describe a hindered rotation about the central site.

Following the above developments, we arrive at the vibronic Hamiltonian [10,11]:

$H_{vibron,z,\varphi} = -(h^2/2I)\Delta(\theta,\varphi) \pm (M\omega^2/G)(D_c-D_b)[r^4(\cos\varphi)^4 + r^4(\sin\varphi)^4 + z^4] \equiv H_{vibron,\varphi} + H_{vibron,z}$

The corresponding complete Schrödinger equation reads

$H_{vibron,z,\varphi} \Psi(z,\varphi) = (E_z + E_\varphi)\Psi(z,\varphi)$ (7)

with eigenvalue $E = E_\varphi + E_z$ and eigenstate $\Psi(z,\varphi) = \Phi(\varphi)A\exp(ikz)$ according to

$H_{vibron,\varphi} \Phi(\varphi) = E_\varphi \Phi(\varphi)$

$H_{vibron,z} \exp(ikz) = E_z \exp(ikz)$ (8)

The angular part has been dealt with in equatorial plane [11]. Its eigenfunctions are Mathieu's periodic functions, while its eigenvalues fall into Mathieu's rotational energy bands [11]. The z-dependent part simplifies to

$H_{vib,z} = -(h^2/2I)(\partial^2/\partial z^2) \pm (M\omega^2/G)(D_c-D_b)z^4 \sim -(h^2/2I)(\partial^2/\partial z^2)$ (z ~ 0) (9)

which is solved as $A\exp(ikz)$ at small z. Its corresponding eigenvalue is given by $E_z = h^2k^2/2I$. At medium and large z the complete Hamiltonian in (7) describes two kinds of anharmonic vibrators, one with a positive curvature, the other one with a negative curvature [3].

The obtained vibronic Schrodinger equation (8) split in two parts shows convincingly that the classical vibronic trajectory is helical [3]. Curiously, this form has been suggested for the light beam propagation [12], despite of the photon being much lighter than the phonon which certainly violates the basic premises of the adiabatic approximation that applies to heavier bosons. In any event, this may have some relationship to photonic crystals.

4.   Antiadiabatic proposal for heavy fermions versus lighter bosons

(examples: electrons vs. photons, nucleons vs. π-mesons, antiadiabatic polarons)

An example is the antiadiabatic polarons: fast local motion but slow translation. Now the boson terms are faster, the electronic terms are parameters, and the complete Hamiltonian is

$H_{comp} = -(h^2/4\pi m)\sum_{i\alpha\beta s}(\partial^2/\partial_i q_{\alpha\beta s}) + \tfrac{1}{2}\sum_{\alpha\beta s}M_{\alpha\beta s}\omega_{\alpha\beta s}^2 q_{\alpha\beta s}^2 + \sum_{n\alpha s} t_{n\alpha s}\, a_{n\alpha s}^\dagger (a_{n-1\alpha s} + a_{n+1\alpha s}) +$

$\sum_{\alpha\beta s}G_{\alpha\beta s}q_{\alpha\beta s}(a_{\alpha s}^\dagger a_{\beta s} + a_{\beta s}^\dagger a_{\alpha s}) + \sum_{\alpha\beta s}E_{\alpha s}a_{\alpha s}^\dagger a_{\alpha s}$

$= -(h^2/4\pi m)\sum_{i\alpha\beta s}(\partial^2/\partial_i q_{\alpha\beta s}) + \tfrac{1}{2}\sum_{\alpha\beta s}M_{\alpha\beta s}\omega_{\alpha\beta s}^2 (q_{\alpha\beta s} + q_{\alpha\beta s 0})^2 + \sum_{n\alpha s} t_{n\alpha s}\, a_{n\alpha s}^\dagger (a_{n-1\alpha s} + a_{n+1\alpha s}) -$

$\tfrac{1}{2}\sum_{\alpha\beta s}\{[G_{\alpha\beta s}(a_{\alpha s}^\dagger a_{\beta s}+a_{\beta s}^\dagger a_{\alpha s})]^2 / M_{\alpha\beta s}\omega_{\alpha\beta s}^2\} + \sum_{\alpha\beta s}E_{\alpha s}a_{\alpha s}^\dagger a_{\alpha s}$, (10)

where the displacement boson coordinate (c-number) reads

$q_{\alpha\beta s 0} = [G_{\alpha\beta s}(a_{\alpha s}^\dagger a_{\beta s}+a_{\beta s}^\dagger a_{\alpha s})] / M_{\alpha\beta s}\omega_{\alpha\beta s}^2$,

with complete eigenvalues and eigenstates

$$E_{comp} = \sum_{n\alpha\beta s} (n + ½)h\nu_{n\alpha s} + \sum_{\alpha\beta s} E_{\alpha s} a_{\alpha s}^\dagger a_{\alpha s} - ½\sum_{\alpha\beta s} \{[G_{\alpha\beta s}(a_{\alpha s}^\dagger a_{\beta s} + a_{\beta s}^\dagger a_{\alpha s})]^2 / M_{\alpha\beta s}\omega_{\alpha\beta s}^2\}$$

$$\Psi(r,q) = \chi(q+q_0)\psi(r), \qquad (11)$$

respectively. Above, we made use of the theory of displaced harmonic oscillators to deriving eigenvalues and eigenstates [13].

We see that there is no point of talking about vibrons and vibronic trajectories in the antiadiabatic case. For antiadiabatic polarons the local configuration is always fully blown before the electron has moved to its next location.

## 5. Brief excerpts of supersymmetry surveys [4,14]

In particle physics, supersymmetry (often abbreviated SUSY) is a symmetry that relates elementary particles of one spin to another particle that differs by half a unit of spin and are known as superpartners. In other words, in a supersymmetric theory, for every type of boson there exists a corresponding type of fermion, and *vice versa* .

As of 2008 there is no direct evidence that supersymmetry is a symmetry of nature. Since superpartners of the particles of the Standard Model have not been observed, supersymmetry, if it exists, must be a broken symmetry allowing the 'sparticles' to be heavy. The minimal supersymmetric Standard Model is one of the best studied candidates for physics beyond the Standard Model.

If supersymmetry exists close to the TeV energy scale, it allows the solution of two major puzzles in particle physics. One is the hierarchy problem - on theoretical grounds there are huge expected corrections to the particles' masses, which without fine-tuning will make them much larger than they are in nature. Another problem is the unification of the weak interactions, the strong interactions and electromagnetism. Another advantage of super-symmetry is that supersymmetric quantum field theory can sometimes be solved. Super-symmetry is a consequence of most versions of string theory, though not a necessary one.

Supersymmetry (SUSY) postulates a deeper relationship between matter particles (spin-1/2 or "fermions") and force carriers (integer spin or "bosons") than the Standard Model (SM). In SUSY, each fermion has a "superpartner" of spin-0 while each boson has a spin-1/2 superpartner. The Higgs sector is also extended to at least five Higgs bosons in the Minimal Supersymmetric Standard Model (MSSM). To this day, no superpartners have been observed: SUSY must be a broken symmetry, i.e. the superpartners must have masses different than those of their partner particles.

Despite the doubling of the spectrum of particles, SUSY has many merits: it is elegant; assuming the existence of superpartners with TeV-scale masses, the Strong, Weak and Electromagnetic force strengths become equal at the same energy of ~ $10^{16}$ GeV (the "GUT scale"); it also provides a natural explanation of why the Higgs mass can be low (< 1 TeV). In SUSY theories, there is even room for explaining the dark matter in the Universe as being due to "neutralinos". If SUSY is a true symmetry of Nature and it is realized at the TeV scale, it will almost certainly be seen in colliders.

## 6. Global (super)symmetry breaking

In the light of the supersymmetry definition, the immediate interpretation of the 'fermion mixing by bosons' symmetry-breaking mechanism (alias the extended Jahn-Teller or vibronic effects) is that it annihilates one superpartner pair, while creating a fermion component of another incipient pair. Care should be taken, however, on applying the supersymmetry terminology, since the two fermions, past and present, should belong to different fermion states or narrow energy bands to effect any mixing.

We note that the details of mixing reflect the fundamental fermionic transitions and the accompanying boson processes of a system of two nearly-degenerate levels (bands). These involve interband fermion transitions along with the emission or absorptions of phonons. The relationship of the mixing Hamiltonian to the supersymmetry partners becomes transparent as one takes account of the spin variable. The spin difference (multiple of ±½ by definition) is taken away by the outgoing fermion:

$$b_{n\alpha +1}^{\dagger} a_{n\alpha +½}^{\dagger} a_{n\beta +½} + b_{n\beta +1}^{\dagger} a_{n\beta +½}^{\dagger} a_{n\alpha +½} + b_{n\beta -1} a_{n\alpha -½}^{\dagger} a_{n\beta -½} + b_{n\alpha -1} a_{n\beta -½}^{\dagger} a_{n\alpha -½} \quad (12)$$

Reading from left to right, a $s=+1$ boson is converted into a $s=+½$ superpartner in one fermion band which leaves a $s=+½$ hole in the other fermion band. Alternatively, a $s=-½$ hole in one fermion band joins the annihilation of a $s=-1$ superpartner boson, while a $s=-½$ unit is created in the other fermion band. This picture is the direct extension of the definitions for the interplay of creation and annihilation operators of fermion and boson superpartners in super-symmetric quantum mechanics through binary products of the form $b_{n-1}^{\dagger} a_{n\alpha -½}^{\dagger}$, etc.[15]. This implies that bosons and fermions are always created in superpartner pairs and the same can be applied to their annihilation, always in pairs too, e.g. $b_{n-1} a_{n\alpha -½}$, etc.

However, unlike superpartners the mixing Hamiltonians of the eqn. (12) type are built up of triple operator products, one phonon and two fermions, which is related to their operating over two fermion bands or levels. In accordance, a superpartner pair $b_{n\alpha +1}^{\dagger} a_{n\alpha +½}^{\dagger}$ composed of a $s=+½$ fermion and a $s=+1$ boson in α-band is created at the expense of the decay of a $s=+½$ fermion in β band which creates both partners. In the third term of eqn. (12) we see another extension of the same principle whereby a superpartner pair is annihilated to create a single $s=-½$ fermion in α band, etc. These events may be considered extending the super-symmetric theory to the two level system.

As a result of fermion mixing, the boson external symmetry changes and it is no longer the higher external symmetry before the mixing event. Eventually, this may come to explain the anticipated supersymmetry lowering or breakup to account for the observed behavior. As stated above, the supersymmetry tells of the internal degrees of freedom which are to combine with the external ones such as the 4D space-time or rotation & reflection transforms.

If fermion mixing destroys a higher external symmetry that is being borne by the annihilated boson, a lower external symmetry will be carried by the created boson. These are the implications of the external symmetry breaking to be rather regarded as symmetry lowering. Then irrespective of the spin partners which may change their position on the external symmetry scale, their status as internal supersymmetry partners may not be affected at all. If this does not suffice, the internal supersymmetry breaking should be looked for elsewhere.

Writing down by opening the brackets in the mixing terms of (1), we note that they all appear in the form of a sum of triple products by partner pairs (created or annihilated) times, on left or on right, the operator of the annihilated or created fermion that seems to have triggered the appearance of the pair, namely,

$(b_{ns}^\dagger + b_{ns})(a_{n\alpha s}^\dagger a_{n\beta s} + a_{n\beta s}^\dagger a_{n\alpha s}) =$

$[b_{ns}^\dagger a_{n\alpha s}^\dagger] a_{n\beta s} + [b_{ns}^\dagger a_{n\beta s}^\dagger] a_{n\alpha s} + a_{n\alpha s}^\dagger [b_{ns} a_{n\beta s}] + a_{n\beta s}^\dagger [b_{ns} a_{n\alpha s}]$ (13)

We have put the s-pair products, created or annihilated, in square brackets that have nothing to do with commutators. Elementary considerations related to two nearly-degenerate bands ($\alpha$ and $\beta$) indicate that the creation of a fermion in the upper band is in parallel with the annihilation of an s-pair in the lower band and, vice versa, the creation of an s-pair in the lower band is in parallel with the annihilation (decay) of a fermion in the upper band. For the decay of particles in supersymmetric theory, see Reference [16].

## 7. Conclusion

In order to arrive at certain conclusions regarding fermions and bosons in general, we have beforehand extended the definitions to upgrade the vibronic effects, defined originally for electron-phonon systems, by making them applicable to a whole of fermion-boson systems and thereby to the appearance and interconversion of superpartner pairs in simple two-level (two-band) fermion systems.

In this endeavor, we pointed out that basic supersymmetry elements, such as the double products of creation operators or double products of annihilation operators of bosons and fermions, respectively, can be found in the mathematical formulation of the Hamiltonian for mixing fermion states by bosons as defined for the vibronic or cooperative Jahn-Teller effects. The vibronic effects come as an universal symmetry-breaking mechanism having deep roots in modern physics and chemistry. If supersymmetry is confirmed experimentally, deciphering its links with the vibronic mixing will become an immediate necessity. All the more so will its relationship to the symmetry-breaking aspects of mixing fermion states by coupling them to bosons.

In pushing our suggestion a bit further, we considered two extreme examples of fermion-boson coupling, known originally from the electron-phonon coupling as well, those of adiabatic and antiadiabatic polarons, respectively. The former one makes it possible to derive a semi-classical trajectory for the polaron under the conditions of a fast fermion motion within a medium that is too slow to accommodate to the fermion displacement. Not surprisingly, the geometric trajectory proved to be that of a helix, as postulated elsewhere for light propagation [12]. Nevertheless, the photon mass being too small, the system of fermions and photons would rather fall in the antiadiabatic regime where no semiclassical trajectory is immediately accountable.